\documentclass[]{proarticle}
\usepackage{multicol}
\usepackage{graphicx}
\usepackage{amssymb}
\usepackage{epsfig}

\def\bea{\begin{eqnarray}}
\def\eea{\end{eqnarray}}
\begin{document}
\title{Finite  BRST-antiBRST Transformations for the Theories with Gauge Group.}
\author{P.Yu. Moshin${}^a$, A.A. Reshetnyak${}^{b,c}$}
\maketitle
\address{${}^a$ Department of Physics, Tomsk State University, Novosobornaya sq., 1, 634050, Tomsk,
Russia, \\
  ${}^b$ Department of Theoretical Physics, Tomsk State Pedagogical University, Kievskaya str., 60, 634061 Tomsk, Russia,\\
 ${}^c$  Institute of Strength Physics and Material Science
SB of RAS,  Akademicheskii av.  2/4, 634021 Tomsk, Russia.}
\eads{moshin@rambler.ru, reshet@tspu.edu.ru}
\begin{abstract}
Following our recent study [P.Yu. Moshin, A.A. Reshetnyak,  Nucl.
Phys. B 888 (2014)  92], we discuss the notion of finite
BRST-antiBRST transformations,
 with a doublet $\lambda_{a}$,
$a=1,2$, of anticommuting (both global and field-dependent)
Grassmann parameters. It turns out that the global finite BRST-antiBRST transformations form
a 2-parametric Abelian supergroup.  We  find an explicit Jacobian corresponding
to this change of variables in the theories with a gauge group.
Special field-dependent BRST-antiBRST transformations  for the
Yang--Mills path integral with $s_{a}$-potential
(functionally-dependent) parameters $\lambda_{a}=s_{a}\Lambda$
generated  by a finite even-valued functional $\Lambda$ and the
anticommuting generators $s_{a}$ of BRST-antiBRST transformations,
amount to a precise change of the gauge-fixing functional. This
proves the independence of the vacuum functional under such
BRST-antiBRST transformations and leads to the presence of
modified Ward identities.  The form of transformation parameters
that induces a change of the gauge in the path integral is found
and is exactly evaluated  for connecting two arbitrary $R_{\xi
}$-like gauges. The finite field-dependent BRST-antiBRST
transformations are used to generalize the Gribov horizon
functional $h_{ 0   }$, in the Landau gauge of the BRST-antiBRST
setting in the Gribov--Zwanziger model, and to find $h_{\xi}$
corresponding to general $R_{\xi}$-like gauges in the form
compatible with a gauge-independent $S$-matrix.
\end{abstract}

%% \MSC 78A35 \sep 78-05 \sep 81V45

\begin{multicols}{2}
\section{Motivations}

The special supersymmetries known as BRST symmetry
\cite{BRST1,BRST2} and BRST-antiBRST symmetry
\cite{aBRST1,aBRST2,aBRST4} provide a basis of the modern
quantization methods for gauge theories \cite{books3,books4}.
These symmetries feature the presence of a
Grassmann-odd parameter $\mu$ and two Grassmann-odd parameters
$(\mu,\bar{\mu})$, respectively. The latter parameters in the
$\mathrm{Sp}\left( 2\right)  $-covariant schemes of generalized
Hamiltonian \cite{BLT1h} and Lagrangian \cite{BLT1,BLT2}
quantization
(see \cite{Hull} as well)  $(\mu,\bar{\mu})\equiv(\mu_{1}%
,\mu_{2})=\mu_{a}$ form an $\mathrm{Sp}\left(  2\right) $-doublet.
These infinitesimal odd-valued parameters may be regarded both as
constants and field-dependent functionals, used, respectively, to
derive the Ward identities and to establish the gauge-independence
of the partition function in the path integral approach.

In  \cite{JM}, the BRST transformations with a finite
field-dependent parameter (FFDBRST) in the gauge theories with a
gauge group, i.e., Yang--Mills (YM) theories, whose quantum action
is constructed by the Faddeev--Popov (FP) rules \cite{FP}, were
first introduced by means of a functional equation for the
parameter in question and used to provide the path integral with
such a change of variables that would allow one to relate the
quantum action given in a certain gauge with the one given in a
different gauge, however, without solving it in the general
setting (see \cite{Upadhyay1} as well).
 The problem of establishing a relation of
the FP action in a certain gauge with the action in a different
gauge by using a change of variables induced by a FFDBRST
transformation was solved in \cite{LL1}, thereby providing an
exact relation between a finite parameter and a finite change of
the gauge-fixing condition in terms of the gauge Fermion. In
particular, this result implies the conservation of the number of
physical degrees of freedom in a given YM theory with respect to
FFDBRST transformations, which means the impossibility of relating
the Yang--Mills theory to another theory, whose action may
contain, in addition to the FP action, certain non-BRST invariant
terms (such as the Gribov horizon functional \cite{Gribov} in the
Gribov--Zwanziger theory \cite{Zwanziger}, as a consequence of
Singer's result \cite{Singer}) in the same configuration
space.\footnote{For the study of the Gribov copies in YM theories,
using the covariant, Landau, maximal Abelian gauges, see
\cite{sorellas, sorellas1, sorellas2, Gongyo, 0806.0348}.}

Notice that the solution of a similar problem for arbitrary
constrained dynamical systems  in the generalized Hamiltonian
formalism \cite{BFV,Henneaux1} has been recently proposed in
\cite{BLThf}, whereas for general gauge theories (possessing
reducible gauge symmetries and/or an open gauge algebra) an exact
Jacobian generated by FFDBRST transformations in the path integral
constructed by the BV procedure \cite{BV} has been obtained in
\cite{Reshetnyak} (see \cite{BLTBV} as well), with a solution of
the consistency of \emph{soft BRST symmetry breaking} \cite{llr1}.

Recently, we have proposed  an extension of BRST-antiBRST
transformations to the case of finite (both global and
field-dependent) parameters in Yang--Mills \cite{MRnew} and
general gauge theories \cite{MRnew2, MRnew3, MRnew1}, using the
Lagrangian and generalized Hamiltonian BRST-antiBRST quantization
methods; also see \cite{BLThfext}. Here, we review the origin of
finite  BRST-antiBRST transformations and use their properties to
study their influence on the quantum structure of YM theoris in
the framework of the BRST-antiBRST setting.

We use the conventions introduced in  \cite{MRnew}. Unless
otherwise specified by an arrow, derivatives with respect to the
fields are taken from the right, and those with respect to the
corresponding antifields are taken from the left. The raising and
lowering of $\mathrm{Sp}\left(  2\right)  $ indices,
$s^{a}=\varepsilon ^{ab}s_{b}$, $s_{a}=\varepsilon_{ab}s^{b}$, is
carried out by a constant antisymmetric metric tensor
$\varepsilon^{ab}$, $\varepsilon
^{ac}\varepsilon_{cb}=\delta_{b}^{a}$, subject to the
normalization $\varepsilon^{12}=1$.

%%%%%%%%%%%%%%%%%%%%%%%%%%%%%%%%%%%%%%
%
\section{Proposal for Finite Field-Dependent BRST-antiBRST Transformations}\label{sec2}
%
%%%%%%%%%%%%%%%%%%%%%%%%%%%%%%%%%%%%%%
The generating functional of Green's functions for irreducible
gauge theories with a closed algebra in the  BRST-antiBRST
Lagrangian quantization \cite{BLT1,BLT2} is given by
 \begin{equation}
Z_F(J)=\textstyle\int d\phi\ \exp\left\{  \frac{i}{\hbar}\left[  S_{F}\left(  \phi\right)
+J_{A}\phi^{A}\right]  \right\}   \label{z(j)}%
\end{equation}
and depends on the sources $J_A$, with the BRST-antiBRST-invariant
quantum action
\begin{eqnarray}
\hspace{-0.5em}&& \hspace{-1em} S_{F}(\phi) = S_{0}\left(  A\right)  + 1/2  s^{a}s_{a}F\left(
A,C\right)\nonumber \\
 \hspace{-0.5em}&& \hspace{-0.5em}=  S_{0}\left(  A\right)  +S_{\mathrm{gf}}\left(  A,B\right)
+S_{\mathrm{gh}}\left(  A,C\right)  +S_{\mathrm{add}}\left(
C\right)
\label{S(A,B,C)}%
\end{eqnarray}
defined in the same total configuration space $\mathcal{M}$ as in
the FP method, and parameterized by the respective classical
fields, the $Sp(2)$-doublet of ghost-antighost fields, and the
Nakanishi--Lautrup fields, $\phi^A =( A^i, C^{\alpha a}, B^\alpha)
$, with the Grassmann parities $\varepsilon(\phi^{A})\equiv
\varepsilon_{A}$, $\varepsilon_{A}  =
\varepsilon(A^i,B^\alpha,C^{\alpha a})\equiv$ $(\varepsilon_i,
\varepsilon_\alpha,\varepsilon_\alpha +1 ) $, using the condensed
notation. The quantities $S_0$, $F$ are the respective classical
gauge-invariant action and an admissible gauge-fixing Bosonic
functional, chosen here in the quadratic approximation for YM
theories, with $A^i=A^{\mu n}(x)$ defined in a $d$-dimensional
Minkowski space and  taking values in the Lie algebra of the
$SU(N)$ gauge group, with  $\eta_{\mu\nu}=diag(-,+,...,+)$,
\begin{equation}\label{YMaction}
  S_0 = - 1/4 \textstyle \int d^d x F_{\mu\nu}^nF^{\mu\nu{}n},  n=1,...,N^2-1,
\end{equation}
in terms of the strength $F^{\mu\nu{}n} = \partial^{[\mu} A^{\nu]
n}+ f^{nop}A^{\mu 0}A^{\nu p}$ and
\begin{equation}\label{gfrxi}
\hspace{-0.7em}F_\xi(  A,C) \hspace{-0.1em} = \hspace{-0.1em} - \frac{1}{2}\textstyle\int d^{d}x\ \left(  A_{\mu}%
^{m}A^{m\mu}-\xi/2\varepsilon_{ab}C^{ma}C^{mb}\right),
\end{equation}
which corresponds to the $R_{\xi}$-family of gauges
with $\chi_\xi(A,B)= \partial _{\mu }A^{\mu a}+\frac{%
\xi }{2}B^{a}=0$ in the FP rules for YM theories.
The remaining terms in (\ref{S(A,B,C)}), i.e., the gauge-fixing term $S_{\mathrm{gf}}$, the ghost term $S_{\mathrm{gh}%
}$, and the interaction term $S_{\mathrm{add}}$, quartic in
$C^{ma}$, are given by
\begin{eqnarray}
\hspace{-0.9em}&&\hspace{-0.9em}S_{\mathrm{gf}}    = \textstyle\int d^{d}x\left[  \left(  \partial^{\mu}A_{\mu
}^{m}\right)  + \xi/2 B^{m}\right]  B^{m},\nonumber \\
\hspace{-0.9em}&&\hspace{-0.9em}S_{\mathrm{gh}}    = \frac{1
}{2}\textstyle\int d^{d}x \left(  \partial^{\mu}C^{ma}\right)  D_{\mu}^{mn}%
C^{nb}\varepsilon_{ab} ,\label{Sadd} \\
\hspace{-0.9em}&&\hspace{-0.9em} S_{\mathrm{add}}    = -\textstyle\frac{\xi}{48}\int d^{d}x\ \ f^{mnl}f^{lrs}%
C^{sa}C^{rc}C^{nb}C^{md}\varepsilon_{ab}\varepsilon_{cd}. \nonumber%
\end{eqnarray}
The action (\ref{YMaction}) is invariant with respect to the
infinitesimal gauge transformations $ \delta A_{\mu}^{m}    =
D_{\mu}^{mn}\zeta^{n}$ with arbitrary functions $\zeta^\alpha
\equiv \zeta^{n}$, $\varepsilon(\alpha)=0$, defied in $R^{1,d-1}$,
whereas the infinitesimal BRST-antiBRST transformations $\delta
\phi^A = s^a\phi^A\mu_{a}$ for YM theories are given in terms of
anticommuting generators $s^a: s^as^ab+s^bs^a=0$,
 \begin{eqnarray}
\hspace{-1.0em}&&\hspace{-0.7em}s^a A_{\mu}^{m}  = D_{\mu}^{mn}C^{na}\ ,\nonumber\\
\hspace{-1.0em}&&\hspace{-0.7em} s^a\delta B^{m}  \hspace{-0.1em} \hspace{-0.1em} =\hspace{-0.1em}  {1}/{2}f^{nml}\hspace{-0.1em}\left( \hspace{-0.1em} B^{l}C^{na}\hspace{-0.1em}+\hspace{-0.1em}({1}/{6})%
f^{lrs}C^{sb}C^{ra}C^{nc}\varepsilon_{cb}\hspace{-0.1em}\right)\hspace{-0.1em},\nonumber\\
\hspace{-1.0em}&&\hspace{-0.7em} s^b C^{ma}  = \left(\hspace{-0.1em}  \varepsilon^{ab}B^{m}-({1}/{2})f^{mnl}%
C^{la}C^{nb}\right)   , \label{DCma}%
\end{eqnarray}
leaving the action  $S_{F}$ and the integrand
$\mathcal{I}^F_{\phi}$ in $Z_F(0)\equiv \int \mathcal{I}^F_{\phi}$
invariant only in the $1$-st order in powers of $\mu_{a}$.

To restore the total BRST-antiBRST invariance
 of $S_{F}$ and $\mathcal{I}^F_{\phi}$ to all orders in $\mu_{a}$,
 we have introduced \cite{MRnew}
finite transformations of  $\phi^{A}$ with a doublet
$\lambda_{a}$ of anticommuting parameters, $\lambda
_{a}\lambda_{b}+\lambda_{b}\lambda_{a}=0$,%
\begin{eqnarray}
\hspace{-1.3em}&& \hspace{-1.1em}\phi^{A}\rightarrow\phi^{\prime A}=\phi^{A}+\Delta\phi^{A}=\phi^{\prime
A}\left(  \phi|\lambda\right) :  \phi^{\prime
}\left(  \phi|0\right)  =\phi , \label{Z}%
\end{eqnarray}
as a solution of the functional equation
\begin{equation}\label{funceq}
  G\left(\phi'\right)  = G\left(\phi\right)\texttt{ if } s^a G\left(\phi\right) = 0
\end{equation}
for any regular  functional $G(\phi)$ invariant
under infinitesimal BRST-antiBRST transformations.
The general solution of (\ref{funceq}) allows one to restore
\emph{finite BRST-antiBRST transformations}
in a unique way $\phi^{A} \to  \phi^{\prime A}$,
\begin{equation}
\phi^{\prime A}\hspace{-0.1em} =\textstyle\hspace{-0.1em}  \phi^{A}\left(1\hspace{-0.1em}+   \overleftarrow{s}^{a}%
 \lambda_{a}\hspace{-0.1em}+\frac{1}{4} \hspace{-0.1em} \overleftarrow{s}^{2}
\lambda^{2}\right)\hspace{-0.1em} \equiv\hspace{-0.1em}
\phi^{A} \exp(\overleftarrow{s}^{a}%
 \lambda_{a} \hspace{-0.1em}), \label{finite1}%
\end{equation}
where the set of elements $\{g(\lambda)\}=\{\exp(\overleftarrow{s}^{a}%
 \lambda_{a} )\}$ forms an Abelian two-parametric supergroup with odd-valued generating elements $\lambda_{a}$.
 The prescription (\ref{Z}), (\ref{funceq})
 for obtaining finite (group) BRST-antiBRST transformations (\ref{finite1})
 works perfectly well in the
  case of changing the configuration space $\mathcal{M}$
  to another representation space in which
  the action of superalgebra of $s^a$ is realized.
  Thus, the finite BRST-antiBRST transformations
  have also been constructed in the generalized
  Hamiltonian \cite{MRnew1} and Lagrangian \cite{MRnew2}
  BRST-antiBRST quantization schemes, which is also
  supported by the Frobenius theorem.
  The BRST-antiBRST invariance of $\mathcal{I}^F_{\phi}$
  implies the validity of the relation
 \begin{equation}\label{finbab}
   \mathcal{I}^F_{\phi g(\lambda)} = \mathcal{I}^F_{\phi},
 \end{equation}
 with allowance for the fact established in \cite{MRnew}
 that the global finite transformations, corresponding to
$\lambda_{a}=\mathrm{const}$, respect the integration measure.

%%%%%%%%%%%%%%%%%%%%%%%%%%%%%%%%
\section{Jacobian of Finite BRST-antiBRST Transformations}\label{sec3}
%
%%%%%%%%%%%%%%%%%%%%%%%%%%%%%%%%
As we have already mentioned, the Jacobian
of the change of variables corresponding
to the global finite transformations is equal to 1:
\begin{equation}
\Im\left(  \phi\right)  =0 \Longrightarrow    \mathrm{Sdet}\left(
\frac{\delta\phi^{\prime}}{\delta\phi}\right)  =1\,\mathrm{and}%
\,d\phi^{\prime}=d\phi. \label{constJ}%
\end{equation}
We have also examined \cite{MRnew}
the finite field-dependent transformations
in the particular case of  functionally-dependent parameters
$\lambda_{a} =\Lambda\overleftarrow{s}_{a}$,
$s^{1}\lambda_{1}+s^{2}\lambda_{2}=-s^{2}\Lambda$,  with a certain
even-valued potential, $\Lambda=\Lambda\left(  \phi\right) $,
inspired by infinitesimal field-dependent BRST-antiBRST
transformations with the parameters
\begin{equation}
\mu_{a}=\frac{i}{2\hbar}\varepsilon_{ab}\left(  \Delta F\right)  _{,A}%
X^{Ab}=\frac{i}{2\hbar}\left(  s_{a}\Delta F\right)  \,, \label{partic_case}%
\end{equation}
which respect the gauge independence of the integrand,
and therefore also of the vacuum functional $Z_{F}(0)$,
with accuracy up  to the terms linear in $\Delta F$:
$\mathcal{I}^F_{\phi g(\mu(\Delta F))}  =
\mathcal{I}^{F+\Delta F}_{\phi} + o(\Delta F)$.
In the  case of finite {field-dependent} transformations
with the group elements $g(\Lambda\overleftarrow{s}_{a})$,
which now form a non-Abelian 2-parametric supergroup,
the superdeterminant of a change of variables takes the form %
\begin{eqnarray}
\hspace{-0.5em}&& \hspace{-1em} \mathrm{Sdet}\left(
\frac{\delta(\phi g(\Lambda\overleftarrow{s}_{a})}{\delta\phi}\right) = \left[  1-\frac{1}{2}%
s^{2}\Lambda\left(  \phi\right)
\right]^{-2},  \label{superJaux}\\
\hspace{-0.5em}&& \hspace{-1em} d\phi^{\prime} = \textstyle  d\phi\ \exp\left\{  \frac{i}{\hbar}\left[
i\hbar\,\mathrm{\ln}\left(  1-\frac{1}{2}s^2\Lambda\right)
^{2}\right]  \right\}  \ . \label{superJ1}%
\end{eqnarray}
%%%%%%%%%%%%%%%%%%%%%%%%%%%%%%%%
\section{Compensation Equation  for  Yang--Mills Theories in Different
Gauges}\label{sec4}
%
%%%%%%%%%%%%%%%%%%%%%%%%%%%%%%%%
In view of the invariance of the quantum action $S_{F}\left(  \phi\right)  $
under (\ref{finite1}), the change $\phi^{A}\rightarrow\phi^{\prime A}=\phi
^{A}g(\lambda(\phi))$ induces in (\ref{z(j)}) the transformation of
the integrand $\mathcal{I}^F_{\phi}$%
\begin{equation}
\hspace{-0.5em}\mathcal{I}^F_{\phi g(\lambda(\phi))}\hspace{-0.3em} = \hspace{-0.1em}\textstyle d\phi \exp\left\{   \frac{i}{\hbar}  \left[
S_{F}\left(  \phi\right)\hspace{-0.1em} + \hspace{-0.1em}i\hbar\,\mathrm{\ln}(  1-\frac{1}{2}s^2\Lambda)
^{2}\right] \hspace{-0.1em} \right\}. \nonumber %
\end{equation}
Due to the explicit form of the initial quantum action $S_{F}=S_{0}-\left(
1/2\right)  F \overleftarrow{s}^2$, the BRST-antiBRST-exact contribution $i\hbar
\,\mathrm{\ln}\left(  1+s^{a}s_{a}\Lambda/2\right)  ^{2}$ to the action
$S_{F}$, resulting from the transformation of the integration measure, can be
interpreted as a change of the gauge-fixing functional made in the original
integrand $\mathcal{I}^F_{\phi}$,%
\begin{eqnarray}
\hspace{-0.7em}&&\hspace{-0.5em}  i\hbar\ \mathrm{\ln}\left(  1+s^{a}s_{a}\Lambda/2\right)  ^{2}%
\ =\ s^{a}s_{a}\left(  \Delta F/2\right)  \label{superJ3m}\\
\hspace{-0.7em}&&\hspace{-0.5em}   \Longrightarrow \mathcal{I}^F_{\phi g(\lambda(\phi))}   =  \hspace{-0.1em} \mathcal{I}^{F+\Delta F}_{\phi}\hspace{-0.1em} , \label{superJ4}%
\end{eqnarray}
for a certain $\Delta F\left(  \phi|\Lambda\right)  $, whose relation to
$\Lambda\left(  \phi\right)  $ is established by (\ref{superJ3m}),
referred to in \cite{MRnew} as
the \emph{compensation equation} for an unknown parameter
$\Lambda(\phi)$, which provides the gauge independence
of the vacuum functional, $Z_F(0)=Z_{F+\Delta F}(0)$.
An explicit solution to (\ref{superJ3m}), which obeys
the solvability condition due to the BRST-antiBRST exactness
of its both sides, is given, up to BRST-antiBRST-exact terms,
by the relations
\begin{eqnarray}
\hspace{-0.7em}&&\hspace{-0.5em}
\Lambda\left(  \phi|\Delta F\right)=\textstyle 2\Delta F\left( \hspace{-0.1em} s^{a}s_{a}\Delta F\right)  ^{-1}\left[ \hspace{-0.2em} \exp\left(\hspace{-0.1em}
\frac{1}{4i\hbar}s^{b}s_{b}\Delta F\right) \hspace{-0.1em} -1\right] \nonumber \\
\hspace{-0.7em}&&\hspace{-0.5em} =\frac{1}{2i\hbar
}\Delta F\sum_{n=0}^{\infty}\frac{1}{\left(  n+1\right)  !}\left(  \frac
{1}{4i\hbar}s^{a}s_{a}\Delta F\right)  ^{n}\ . \label{Lambda-Fsol1}%
\end{eqnarray}
Conversely, having considered the equation (\ref{superJ3m})
for an unknown $\Delta F$ with a given $\Lambda$, we obtain
\begin{eqnarray}\hspace{-0.7em}&&\hspace{-0.5em}
\Delta F\left(  \phi\right)  =\textstyle- 2i\hbar\ \Lambda \left(
s^2\Lambda  \right)  ^{-1}\ln\left(  1-s^2%
\Lambda  /2\right)  ^{2}\ . \label{Lambda-Fsol}%
\end{eqnarray}
Thus, the
field-dependent transformations  with the parameters $\lambda_{a}=s_{a}\Lambda$
amount to a precise change of the gauge-fixing functional.
For instance, in order to relate $Z_{F_\xi}(J)$ to
$Z_{F_{\xi+\Delta \xi }}(J)$ in the $R_\xi$-family of gauges,
one has to carry out FFDBRST-antiBRST transformations with the parameters
\begin{eqnarray}
\hspace{-0.9em}&&\hspace{-0.7em}\lambda_{a} \textstyle =\frac{\Delta\xi}{4i\hbar}\varepsilon_{ab}\int
d^{d}x\ \left( B^{n}C^{nb}\sum\limits_{n=0}^{\infty}\frac{1}{\left(  n+1\right)  !}
%+\frac{1}{2}f^{nml}C^{lc}C^{mb}C^{nd}\varepsilon_{cd}
\right)\left[
\frac
{\Delta\xi}{4i\hbar}\int d^{d}y\right.
\nonumber\\
\hspace{-0.9em}&&\hspace{-0.7em}\times
\left. \hspace{-0.3em}\left(  \hspace{-0.2em}B^{u}B^{u}-\frac{1}{24}%
f^{uwt}f^{trs}C^{sc}C^{rp}C^{wd}C^{uq}\varepsilon_{cd}\varepsilon
_{pq}\hspace{-0.1em}\right)  \hspace{-0.3em}\right]\hspace{-0.2em} ^{n}\hspace{-0.4em}. \label{lamaxi}%
\end{eqnarray}
%%%%%%%%%%%%%%%%%%%%%%%%%%%%%%%%
\section{Gauge Dependence Problem and modified Ward identities}\label{sec5}
%
%%%%%%%%%%%%%%%%%%%%%%%%%%%%%%%%
In \cite{MRnew3}, the property (\ref{superJ4}) leads
to a so-called modified Ward identity,
depending on field-dependent parameters,
$\lambda_{a} =\Lambda\overleftarrow{s}_{a}$, and
therefore also on a finite change of the gauge, in view of (\ref{Lambda-Fsol1}),
\begin{eqnarray}
\hspace{-1.3em}&&\hspace{-0.9em} \textstyle\left\langle \left\{  1+\frac{i}{\hbar}J_{A}\left[ X^{Aa} \lambda_{a}%
(\Lambda)-\frac{1}{2}Y^{A}\lambda^{2}(\Lambda)\right]-\frac{1}{4}\left(
\frac{i}{\hbar}\right)  {}^{2}\varepsilon_{ab}  \right.\right.\nonumber \\
\hspace{-1.3em}&&\hspace{-0.9em} \textstyle\left.\left. \times J_{A}X^{Aa} J_{B}X^{Bb}%
\lambda^{2}(\Lambda)\right\}  \left(  1-\frac{1}{2}\Lambda\overleftarrow
{s}^{2}\right)  {}^{-2}\right\rangle _{F,J} =1 , \label{mWIclalg}%
\end{eqnarray}
for $ (s^a\phi^A, s^2\phi^A) \equiv (X^{Aa},-2 Y^A)$.
The property (\ref{superJ4}) also implies a relation describing the gauge
dependence of $Z_{F}(J)$ for a finite change of the gauge $F\rightarrow
F+\Delta F$:%
\begin{eqnarray}
\hspace{-0.9em}&&\hspace{-0.5em} \Delta Z_{F}(J)\textstyle =Z_{F}(J)\left\langle \frac{i}{\hbar}%
J_{A}\left[ X^{Aa} \lambda_{a}\left(  \phi|-\Delta{F}\right)   \right.\right.
\nonumber\\
\hspace{-0.9em}&&\hspace{-0.5em} \textstyle  \left.-\frac{1}%
{2}Y^{A}\lambda^{2}\left(  \phi|-\Delta{F}\right)  \right] - \left.  (-1)^{\varepsilon_{B}}\left(  \frac{i}{2\hbar}\right)  ^{2}%
J_{B}J_{A}\right.
\nonumber\\
\hspace{-0.9em}&&\hspace{-0.5em} \textstyle  \left. \times\left(  X^{Aa}X^{Bb}\right)  \varepsilon_{ab} \lambda^{2}\left(
\phi|-\Delta{F}\right) \right\rangle _{F,J} , \label{GDInew1}%
\end{eqnarray}
which means the on-shell ($J=0$) gauge independence of the conventional
$S$-matrix due to equivalence theorem \cite{equiv}.
In (\ref{mWIclalg}), (\ref{GDInew1}),
the symbol \textquotedblleft$\langle\mathcal{A}\rangle_{F,J}%
$\textquotedblright\ for a quantity $\mathcal{A}=\mathcal{A}(\phi)$
denotes the source-dependent average expectation value for a
gauge-fixing $F(\phi)$
\begin{equation}\nonumber
\left\langle \mathcal{A}\right\rangle _{F,J}=\textstyle Z_{F}^{-1}(J)\int d\phi
\ \mathcal{A}\left(  \phi\right)  \exp\left\{  \frac{i}{\hbar}\left[
{S}_{F}  +J\phi\right]  \right\},
\end{equation}
with $\left\langle 1\right\rangle _{F,J}=1$.
For constant $\lambda_{a}$, the relation (\ref{mWIclalg}) implies
an $\mathrm{Sp}(2)$-doublet of the usual Ward identities at the first order in
$\lambda_{a}$: $J_{A}\left\langle X^{Aa}\right\rangle _{F,J}=0$  and a derivative identity at the second
order in $\lambda_{a}$:
\begin{equation}
  \left\langle J_{A}\left[ 2Y^{A}
+\left(  i/\hbar\right)\varepsilon_{ab} X^{Aa}  J_{B} X^{Bb}
\right]  \right\rangle _{F,J}=0\ . \label{WIlag3}%
\end{equation}

\section{Gribov--Zwanziger Theory in BRST-antiBRST Formulation of Landau and
Feynman Gauges}\label{sec6}
%%%%%%%%%%%%%%%%%%%%%%%%%%%%%%%%
Since the gauge-fixing functional $F_0$ corresponds
to the Landau gauge, we introduce the Gribov horizon
functional in the same manner as in \cite{Zwanziger}
for the FP procedure in the Euclidian space:
\begin{eqnarray}
\hspace{-0.9em}&&\hspace{-0.5em}
\label{gribovH}h(  A)  =\int d^{d}x\ d^{d}y\ f^{mrl}%
A_{\mu}^{r}(  x)    {K ^{mn}}^{-1}(  x;y) \nonumber \\
\hspace{-0.9em}&&\hspace{-0.5em}
f^{nsl}A^{\mu{}s }(  y)  + d(  N^{2}-1),
\end{eqnarray}
with $\gamma^{2}$, $ K^{mn}(  x;y)$
being the Gribov mass parameter determined from
the gap equation and the FP matrix.
A proper BRST-antiBRST-non-invariant action
in the $F_0$ reference frame has the form
\begin{equation}
\label{GZbab}S_{h}\left(  \phi\right)  =S_{F_{0}}\left(  \phi\right)
+\gamma^{2}h\left(  \phi\right)  \,.
\end{equation}
We determine the Gribov--Zwanziger theory
in any $F_\xi$ gauges ($R_\xi$-gauges) in a way
compatible with the gauge-independence of the
generating functional of Green's functions in $F_0$,
where Gribov horizon in the gauge $F_\xi$
should be determined as
 \begin{eqnarray}
\hspace{-0.9em}&&\hspace{-0.5em} h_{\xi}    =h\Big(1+\frac{1}{2i\hbar}\left(  \overleftarrow{s}^{a}\right)  \left(  \Delta
F_{\xi}\overleftarrow{s}_{a}  \right) \sum_{n=0}^{\infty}\frac{1}{\left(
n+1\right)  !}\nonumber\\
\hspace{-0.9em}&&\hspace{-0.5em} \times \left(-  \frac{1}{4i\hbar}\Delta F_{ \xi}\overleftarrow{s}^2
\right)  ^{n}  -\frac{1}{16\hbar^{2}}\left(  \overleftarrow{s}^{2}\right)  \left(  \Delta F_{ \xi}  \right)  ^{2}\nonumber\\
\hspace{-0.9em}&&\hspace{-0.0em} \times \left[  \sum_{n=0}^{\infty}\frac{1}{\left(
n+1\right)  !}\left( - \frac{1}{4i\hbar}\Delta F_{ \xi}\overleftarrow{s}^2
\right)  ^{n}\right]  ^{2}\Big) , \label{gribovHxi}%
\end{eqnarray}
where $\Delta F_{ \xi}$ is readily determined with account taken
of (\ref{lamaxi}); for details, see \cite{MRnew}.
The construction of the Gribov horizon functional $h_{\xi}$
in the gauge $F_\xi$, starting from $h_{0}=h(  A)$ in the gauge $F_0$,
may be considered as a generalization of the result \cite{LL2}
obtained in the BRST setting of the problem.
Considering, instead of $h_{0}$, a functional
calculated, e.g., in the Coulomb gauge \cite{HFZwanziger},
we may use the above recipe to construct the form
of $h_{\xi}$ in any $F_{ \xi}$-gauges,
including the Landau and Feynman gauges,
and compare them with those obtained by
the non-perturbative Zwanziger's recipe \cite{Zwanziger},
in order to verify the consistency of Zwanziger's
recipe with the requirement of gauge independence
for the corresponding  vacuum functionals. 

\section{Conclusion}
%%%%%%%%%%%%%%%%%%%%%%%%%%%%%%%%
We have proposed the concept of finite BRST-antiBRST and FFDBRST-antiBRST
transformations for Yang--Mills theories in the $\mathrm{Sp}(2)$-covariant
Lagrangian quantization.
The Jacobian of the change of variables generated
by FFDBRST-antiBRST transformations
with functionally dependent parameters is exactly calculated.
It is established that quantum YM actions
in different gauges are related to each other
by means of FFDBRST-antiBRST transformations
with functionally dependent parameters obtained
as solutions of the compensation equation.
A new Ward identity and the gauge dependence
problem for finite changes of the gauge
for the  generating functional of Green's functions
are derived and studied.
The Gribov-Zwanziger theory
in a BRST-antiBRST formulation
is suggested and the Gribov horizon functional, $h$,
in the Feynman and arbitrary gauges, starting from $h_0$ given in Landau gauge, is suggested
in a way compatible with the gauge independence of
the $S$-matrix. Concluding, note that Gribov's problem for YM theories in BRST-antiBRST setting may be elaborated with help of composite fields technique following to the results  \cite{Reshetnyak2}  obtained within BRST setting of the problem.

%%%%%%%%%%%%%%%%%%%%%%%%%%%%%%%%
\section*{Acknowledgement}
The study has been supported by the RFBR grant, Project
No. 12-02-00121, and by the grant for LRSS, Project No. 88.2014.2.
It has also been partially supported by the Ministry of
Science of the Russian Federation, Grant No. 2014/223.

\end{multicols}
%%%%%%%%%%%%%%%%%%%%%%%%%%%%%%%%

\end{document}